\documentclass[english]{article}
\usepackage{mathptmx}
\usepackage[T1]{fontenc}
\usepackage[latin9]{inputenc}
\usepackage{geometry}
\usepackage{subfigure}
\usepackage{amsmath}
\usepackage{color}
\usepackage{graphicx}
\usepackage{amssymb}

\makeatletter

\providecommand{\tabularnewline}{\\}

\newcommand{\lyxaddress}[1]{
\par {\raggedright #1
\vspace{1.4em}
\noindent\par}
}

\usepackage{axodraw}

\usepackage{babel}
\makeatother

\begin{document}
\newcommand{\gev}{\ensuremath{\mathrm{\, Ge\kern-0.1em V}}}
\newcommand{\qq}{\ensuremath{\, q_{4}\bar{q_{4}}}}

\newcommand{\ptleptcut}{\ensuremath{\, p_{T}^{lept}>15\gev}}
\newcommand{\ptjetcut}{\ensuremath{\, p_{T}^{jet}>20\gev}}

\newcommand{\hardjetcut}{\ensuremath{\, p_{T}^{jet}>100\gev}}

\newcommand{\mWhad}{\ensuremath{\, m_{jj}^{W}}}

\title{Search for 4th family quarks with the ATLAS detector}

\author{V. E. \"Ozcan$^{1}$ , S. Sultansoy$^{2,3}$ and G. \"Unel$^{4,5}$}

\maketitle

\lyxaddress{$^{1}$ Department of Physics and Astronomy, University College London,
London, UK\textcolor{black}{.}\\
\textcolor{black}{$^{2}$ Institute of Physics, Academy of Sciences,
Baku, Azerbaijan. }\\
\textcolor{black}{$^{3}$ TOBB ETU University, Physics Department,
Ankara, Turkey.}\\
\textcolor{black}{$^{4}$ CERN, Physics Department, Geneva, Switzerland.
}\\
\textcolor{black}{5 University of California at Irvine, Physics Department,
USA. }}

\begin{abstract}
The pair production of heavy fourth-generation quarks, which are predicted
under the hypothesis of flavor democracy, is studied using tree-level
Monte Carlo generators and fast detector simulation. Two heavy-quark
mass values, 500 and 750$\gev$, are considered with the assumption
that the fourth family mixes primarily with the two light families.
It is shown that a clear signature will be observed in the data collected
by the ATLAS detector, after the first year of low-luminosity running
at the Large Hadron Collider.
\end{abstract}

\section{Introduction}

It is well known that the number of fundamental fermion families (generations)
is not fixed by the Standard Model (SM). The precision measurements
performed by the Large Electron-Positron Collider experiments at the
$Z$ pole have shown that the number of families with light neutrinos
($m_{\nu}<m_{Z}/2$ ) is equal to three. On the other hand, the asymptotic
freedom in QCD constrains this number to be less than nine. Therefore,
from a pure experimentalist approach, it is meaningful to search for
a possible fourth SM family at the forthcoming colliders. On the theoretical
side, the fourth SM family is a direct outcome of the flavor democracy
(or in other words democratic mass matrix) approach \cite{DMM,DMM-2,DMM-3}
which is strongly motivated by the naturalness arguments (see the
review \cite{FLDem review} and the references therein). Meanwhile,
there are phenomenological arguments against the existence of a fifth
SM family \cite{no-fifth}. In this paper, the additional quark and
lepton pairs of the fourth family are denoted as $u_{4}$, $d_{4}$
and $e_{4}$, $\nu_{4}$. 

The most recent limit on the mass of the $u_{4}$ quark is $m_{u_{4}}>256$
$\gev$ \cite{R-CDF-t'}. The partial wave unitarity gives an upper
bound of about 1 TeV to the fourth family fermion masses \cite{partial-wave-unitarity}.
According to flavor democracy, the masses of the new quarks have to
be within few GeV of each other. This is also experimentally hinted
by the value of the $\rho$ parameter which is close to unity \cite{PDG}.
Therefore, if the fourth SM family exists, the Large Hadron Collider
(LHC) will copiously produce its quarks \cite{R-atlas-tdr} and the
proposed linear colliders will provide opportunity to discover its
leptons \cite{4th-linearcollider}. As the single production of the
new quarks in LHC is suppressed as compared to their pair production,
due to the small value of the CKM matrix elements, the latter is considered.
The new quarks, being heavy, will decay to the known SM quarks and
$W$ bosons. The dominant decay channels are defined by the $4\times4$
extension of the CKM mixing matrix with two distinct possibilities: 

1) If the fourth family is primarily mixing with the third one, the
decay channels will be $u_{4}\rightarrow W^{+}b$ and $d_{4}\rightarrow W^{-}t$.
The signature of the $u_{4}\,\bar{u_{4}}$ production will be $W^{+}W^{-}b\,\bar{b}$
whereas in the case of $d_{4}\,\bar{d_{4}}$, the final state would
have an additional $W^{+}W^{-}$ pair. The former case has been studied
in \cite{R-atlas-tdr,4th-fam-to-3rd} about 10 years ago$\,$%
\footnote{Recently this process has been reconsidered in \cite{Holdom} as {}``the
best scenario for the LHC''.%
}. The latter case, while potentially feasible owing to the low predicted
SM backgrounds with four $W$ bosons in the final state, is likely
to be less interesting as a discovery channel, due to the difficulties
in the jet association and invariant mass reconstruction. 

2) If the fourth generation is primarily mixing with the first two
families, the dominant decay channels will be $u_{4}\rightarrow W^{+}d/s$
and $d_{4}\rightarrow W^{-}u/c$. In this case, since the light quark
jets are indistinguishable, the signature will be $W^{+}W^{-}j\, j$
for both $u_{4}\,\bar{u}_{4}$ and $d_{4}\,\bar{d}_{4}$ pair production.
Therefore, both up and down type new quarks should be considered together
since distinguishing between $u_{4}$ and $d_{4}$ quarks with quasi-degenerate
masses at hadron collider seems to be a difficult task. In this sense,
lepton colliders are more advantageous, especially if the fourth family
quarkonia could be formed.

Results of the most up-to-date measurements on the quark mixings as
published by the Particle Data Group \cite{PDG} together with the
unitarity assumption of the $4\times4$ extension of the CKM matrix
can be used to constrain the fourth-family quark related mixings.
The first step is to calculate the squares of the entries in the fourth
row and column together with their errors: 

\begin{eqnarray}
V_{i4}^{2} & = & 1-\sum_{j=1}^{3}V_{ij}^{2}\\
V_{4i}^{2} & = & 1-\sum_{j=1}^{3}V_{ji}^{2}\nonumber \\
\delta V_{4i}^{2} & = & \sqrt{4\times\sum_{j=1}^{3}(V_{ji}\times\delta_{ji})^{2}}\nonumber \\
\delta V_{i4}^{2} & = & \sqrt{4\times\sum_{j=1}^{3}(V_{ij}\times\delta_{ij})^{2}}\nonumber \end{eqnarray}
where $V_{ij}$ are the CKM matrix elements and the $\delta_{ij}$
are the quoted errors on these measurements. If one allows the $V_{i4}^{2}$
and $V_{4i}^{2}$ to deviate by one sigma, the square root of the
sum gives the upper limit for the fourth family quark mixings:

\begin{eqnarray}
\hbox{CKM}{}_{4\times4} & = & \left[\,\begin{array}{cccc}
0.97377\pm0.00027 & 0.2257\pm0.0021 & 0.00431\pm0.00030 & <0.044\\
0.230\pm0.011 & 0.957\pm0.095 & 0.0416\pm0.0006 & <0.46\\
0.0074\pm0.0008 & 0.0406\pm0.0027 & >0.78 & <0.47\\
<0.063 & <0.46 & <0.47 & >0.57\end{array}\,\right]\end{eqnarray}

where the lower (upper) limit of 0 (1) is implicitly assumed for all
the new entries\cite{Tait}. 

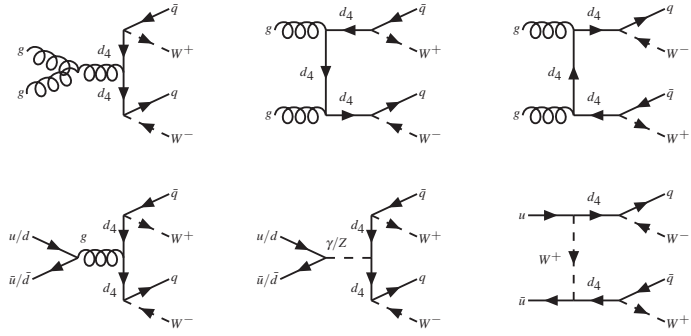
\begin{figure}
{
\unitlength=1.0 pt
\SetScale{1.0}
\SetWidth{0.7}      
\tiny    
{} \qquad\allowbreak
\begin{center}
\begin{picture}(79,81)(0,0)
\Text(11.0,57.0)[r]{$g$}
\Gluon(12.0,57.0)(31.0,49.0){3.0}{3} 
\Text(11.0,41.0)[r]{$g$}
\Gluon(12.0,41.0)(31.0,49.0){3.0}{3} 
\Gluon(31.0,49.0)(48.0,49.0){3.0}{3}
\Text(44.0,57.0)[r]{$d_4$}
\ArrowLine(48.0,65.0)(48.0,49.0) 
\Text(66.0,73.0)[l]{$\bar{q}$}
\ArrowLine(65.0,73.0)(48.0,65.0) 
\Text(66.0,57.0)[l]{$W^+$}
\DashArrowLine(48.0,65.0)(65.0,57.0){3.0} 
\Text(44.0,41.0)[r]{$d_4$}
\ArrowLine(48.0,49.0)(48.0,33.0) 
\Text(66.0,41.0)[l]{$q$}
\ArrowLine(48.0,33.0)(65.0,41.0) 
\Text(66.0,25.0)[l]{$W^-$}
\DashArrowLine(65.0,25.0)(48.0,33.0){3.0} 
\end{picture} \ 
{} \qquad\allowbreak
\begin{picture}(79,81)(0,0)
\Text(11.0,65.0)[r]{$g$}
\Gluon(12.0,65.0)(31.0,65.0){3.0}{3.0} 
\Text(39.0,69.0)[b]{$d_4$}
\ArrowLine(48.0,65.0)(31.0,65.0) 
\Text(66.0,73.0)[l]{$\bar{q}$}
\ArrowLine(65.0,73.0)(48.0,65.0) 
\Text(66.0,57.0)[l]{$W^+$}
\DashArrowLine(48.0,65.0)(65.0,57.0){3.0} 
\Text(27.0,49.0)[r]{$d_4$}
\ArrowLine(31.0,65.0)(31.0,33.0) 
\Text(11.0,33.0)[r]{$g$}
\Gluon(12.0,33.0)(31.0,33.0){3.0}{3.0} 
\Text(39.0,37.0)[b]{$d_4$}
\ArrowLine(31.0,33.0)(48.0,33.0) 
\Text(66.0,41.0)[l]{$q$}
\ArrowLine(48.0,33.0)(65.0,41.0) 
\Text(66.0,25.0)[l]{$W^-$}
\DashArrowLine(65.0,25.0)(48.0,33.0){3.0} 
\end{picture} \ 
{} \qquad\allowbreak
\begin{picture}(79,81)(0,0)
\Text(11.0,65.0)[r]{$g$}
\Gluon(12.0,65.0)(31.0,65.0){3.0}{3}
\Text(39.0,69.0)[b]{$d_4$}
\ArrowLine(31.0,65.0)(48.0,65.0) 
\Text(66.0,73.0)[l]{$q$}
\ArrowLine(48.0,65.0)(65.0,73.0) 
\Text(66.0,57.0)[l]{$W^-$}
\DashArrowLine(65.0,57.0)(48.0,65.0){3.0} 
\Text(27.0,49.0)[r]{$d_4$}
\ArrowLine(31.0,33.0)(31.0,65.0) 
\Text(11.0,33.0)[r]{$g$}
\Gluon(12.0,33.0)(31.0,33.0){3.0}{3} 
\Text(39.0,37.0)[b]{$d_4$}
\ArrowLine(48.0,33.0)(31.0,33.0) 
\Text(66.0,41.0)[l]{$\bar{q}$}
\ArrowLine(65.0,41.0)(48.0,33.0) 
\Text(66.0,25.0)[l]{$W^+$}
\DashArrowLine(48.0,33.0)(65.0,25.0){3.0} 
\end{picture} \ 
\end{center}
}
\vspace{-1.2cm}

{
\unitlength=1.0 pt
\SetScale{1.0}
\SetWidth{0.7}      
\tiny    
{} \qquad\allowbreak
\begin{center}
\begin{picture}(79,81)(0,0)
\Text(13.0,57.0)[r]{$u/d$}
\ArrowLine(14.0,57.0)(31.0,49.0) 
\Text(13.0,41.0)[r]{$\bar{u}/\bar{d}$}
\ArrowLine(31.0,49.0)(14.0,41.0) 
\Text(32.0,55.0)[lb]{$g$}
\Gluon(31.0,49.0)(48.0,49.0){3.0}{3.0} 
\Text(46.0,61.0)[r]{$d_4$}
\ArrowLine(48.0,65.0)(48.0,49.0) 
\Text(66.0,73.0)[l]{$\bar{q}$}
\ArrowLine(65.0,73.0)(48.0,65.0) 
\Text(66.0,57.0)[l]{$W^+$}
\DashArrowLine(48.0,65.0)(65.0,57.0){3.0} 
\Text(46.0,37.0)[r]{$d_4$}
\ArrowLine(48.0,49.0)(48.0,33.0) 
\Text(66.0,41.0)[l]{$q$}
\ArrowLine(48.0,33.0)(65.0,41.0) 
\Text(66.0,25.0)[l]{$W^-$}
\DashArrowLine(65.0,25.0)(48.0,33.0){3.0} 
\end{picture} \ 
{} \qquad\allowbreak
\begin{picture}(79,81)(0,0)
\Text(13.0,57.0)[r]{$u/d$}
\ArrowLine(14.0,57.0)(31.0,49.0) 
\Text(13.0,41.0)[r]{$\bar{u}/\bar{d}$}
\ArrowLine(31.0,49.0)(14.0,41.0) 
\Text(31.0,52.0)[lb]{$\gamma/Z$}
\DashLine(31.0,49.0)(48.0,49.0){3.0} 
\Text(46.0,61.0)[r]{$d_4$}
\ArrowLine(48.0,65.0)(48.0,49.0) 
\Text(66.0,73.0)[l]{$\bar{q}$}
\ArrowLine(65.0,73.0)(48.0,65.0) 
\Text(66.0,57.0)[l]{$W^+$}
\DashArrowLine(48.0,65.0)(65.0,57.0){3.0} 
\Text(46.0,37.0)[r]{$d_4$}
\ArrowLine(48.0,49.0)(48.0,33.0) 
\Text(66.0,41.0)[l]{$q$}
\ArrowLine(48.0,33.0)(65.0,41.0) 
\Text(66.0,25.0)[l]{$W^-$}
\DashArrowLine(65.0,25.0)(48.0,33.0){3.0} 
\end{picture} \ 
{} \qquad\allowbreak
\begin{picture}(79,81)(0,0)
\Text(13.0,65.0)[r]{$u$}
\ArrowLine(14.0,65.0)(31.0,65.0) 
\Text(39.0,69.0)[b]{$d_4$}
\ArrowLine(31.0,65.0)(48.0,65.0) 
\Text(66.0,73.0)[l]{$q$}
\ArrowLine(48.0,65.0)(65.0,73.0) 
\Text(66.0,57.0)[l]{$W^-$}
\DashArrowLine(65.0,57.0)(48.0,65.0){3.0} 
\Text(27.0,49.0)[r]{$W^+$}
\DashArrowLine(31.0,65.0)(31.0,33.0){3.0} 
\Text(13.0,33.0)[r]{$\bar{u}$}
\ArrowLine(31.0,33.0)(14.0,33.0) 
\Text(39.0,37.0)[b]{$d_4$}
\ArrowLine(48.0,33.0)(31.0,33.0) 
\Text(66.0,41.0)[l]{$\bar{q}$}
\ArrowLine(65.0,41.0)(48.0,33.0) 
\Text(66.0,25.0)[l]{$W^+$}
\DashArrowLine(48.0,33.0)(65.0,25.0){3.0} 
\end{picture} \ 
\end{center}
}
\label{fig:The-three-level}\vspace{-0.9cm}

\caption{The tree-level Feynman diagrams for the pair production and decay
of the $d_{4}$ quarks at the LHC.}

\end{figure}

The remaining of this paper investigates the discovery potential of
ATLAS experiment at the LHC accelerator for the fourth family quarks
in the case where their dominant mixings are to first and second SM
families as described in the second scenario above. The tree level
diagrams for the pair production of the new quarks and their subsequent
decays are given in Fig. \ref{fig:The-three-level} for the $d_{4}$
quark decaying via $d_{4}\to W\, q\quad(q=u,\, c)$ . The same diagrams
are also valid for the $u_{4}$ quark production and decay, provided
$c$ and $u$ quarks are replaced by $s$ and $d$ quarks. The widths
of the $d_{4}$ and $u_{4}$ quarks are proportional to $|V_{d_{4}u}|^{2}+|V_{d_{4}c}|^{2}$
and $|V_{u_{4}d}|^{2}+|V_{u_{4}s}|^{2}$ respectively. Although the
extension parameters have much higher upper limits, for the event
generation and analysis section, the common and conservative value
of 0.01 is used for all four relevant mixings. As the widths of the
new quarks are much smaller than their masses, this selection of the
new CKM elements has no impact on the pair production cross sections. 

\section{Event Generation}

In order to study the possibility of discovery, the four-family model
has been implemented into the tree-level generator, CompHEP v4.4.3\cite{R-comphep}
and the pair production of the new quarks at the LHC and their subsequent
decay into SM particles have been simulated. The QCD scale is set
to the mass of the new quark under study and CTEQ6L1 set is chosen
for the parton distribution functions \cite{R-cteq}. Table \ref{tab:The-considered-masses}
gives the cross section for the $d_{4}\,\bar{d_{4}}$ production for
three example values of $d_{4}$ quark mass together with the decay
widths. As the cross section for $u_{4}\,\bar{u_{4}}$ production
is within 1\% of the $d_{4}\,\bar{d_{4}}$ one, from this point on
only $d_{4}$ will be considered and the results will be multiplied
by two to cover all signal processes involving both $u_{4}$ and $d_{4}$
quarks. For each of the considered mass values, 12 thousand signal
events have been generated for the $d_{4}\,\bar{{d_{4}}}\rightarrow W^{-}\, W^{+}\, j\, j$
process where $j$ is a jet originating from a quark or antiquark
of the first two SM families. To benefit from the possible lepton
and jet combined triggers and to reduce the ambiguity in the invariant
mass reconstruction, the hadronic decay of one $W$ boson and the
leptonic (electron or muon) decays of the other one have been considered.
Therefore, the signal is searched for in the $4j+\ell+E\!\!\!/_{T}$
final state where $\ell$ is an electron or a muon. 

The backgrounds events originate from all the SM processes whose final
state has at least two $W$ bosons and two non b-tagged jets. The
direct background is from SM events which yield exactly the same final
state particles as the signal events. The contributions from same
sign $W$ bosons are insignificant. Some of the indirect backgrounds
are also taken into account. The dominant contribution is from $t\,\bar{t}$
pair production where the $b$ jets from the decay of the top quark
could be mistagged as a light jet. Similarly the jet associated top-quark
pair production ( $t\,\bar{{t}}\, j\rightarrow W^{-}\, W^{+}\, b\,\bar{b}\, j$
) contributes substantially to the SM background as the production
cross section is comparable to the pair production and only one mistagged
$b$-jet is sufficient to fake the signal events. The cross section
for the next-order process, namely $p\, p\to t\,\bar{t}\,2j$, has
been computed to be four times smaller than $t\,\bar{t}\: j$ process
and therefore this process has not been considered. It should be noted
that the $t\,\bar{t}$ and $t\,\bar{{t}}\, j$ samples have been conservatively
added together, in spite of the fact that initial and final-state
parton showers simulated in Pythia for the former would account for
part of the cross section for the latter. Finally, background from
SM processes with $W^{\pm}\, Z\, q\,\bar{q}\,\,\,(q=u,d,s,c)$ final
state has been studied. Its contribution to the total background is
very similar to the direct ($WWjj$) background. All the mentioned
background processes have been generated with MadGraph v3.95\cite{R-madgraph}.
This tree-level generator was previously shown to give results in
good agreement with CompHEP and to be more suitable for running on
a computer farm\cite{E6-ilk}. A total of more than 280 thousand events
generated at different QCD scales and jet selection criteria comprise
the background sample.

The events from both generators are fed into the ATLAS detector simulation
and event reconstruction framework, ATHENA v11.0.41, with the CompHEP
events using the interface program CPYTH v2.0.1 \cite{R-cpyth}. Parton
showering, hadronization and fragmentation are simulated using the
ATHENA interface of Pythia v6.23 \cite{R-pythia} and the detector
response is obtained from the fast simulation software, ATLFast\cite{R-AtlFast}.
This software uses a parameterized function to calculate the final
particle kinematic variables rapidly, and its output is calibrated
to match the results from GEANT-based full detector simulation\cite{Geant}.
The physics objects from ATLFast are used in the final analysis in
ROOT 5.12 \cite{R-root}.

\begin{table}
\caption{The quark-mass values considered and the associated width and pair
production cross sections at the LHC.}

\begin{centering}
\begin{tabular}{c|c|c}
$m_{d_{4}}$(GeV) & 500 & 750\tabularnewline
\hline
\hline 
$\Gamma$(GeV) & 8.23$\times10^{-3}$ & 2.79$\times10^{-2}$\tabularnewline
\hline 
$\sigma$(pb) & $2.63$ & $0.250$\tabularnewline
\end{tabular}\label{tab:The-considered-masses}
\par\end{centering}
\end{table}

\section{Event Selection and Reconstruction}

\begin{figure}
\begin{centering}
\subfigure[]{\includegraphics[width=0.42\textwidth,height=4.8cm]{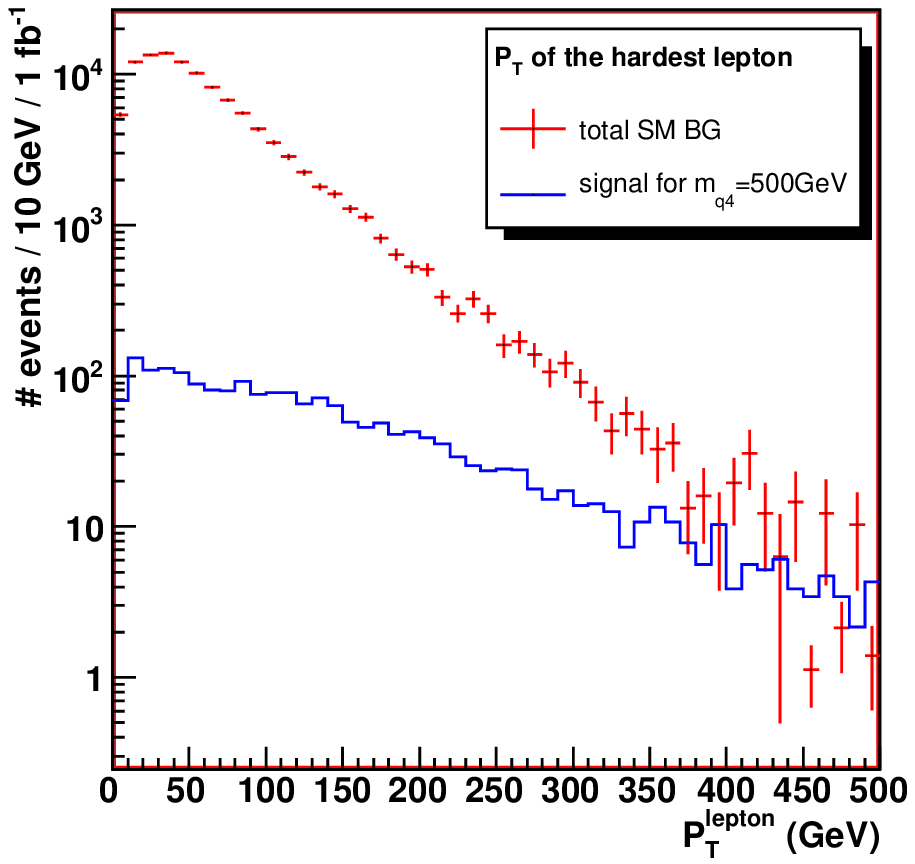}}\subfigure[]{\includegraphics[width=0.42\textwidth,height=4.8cm]{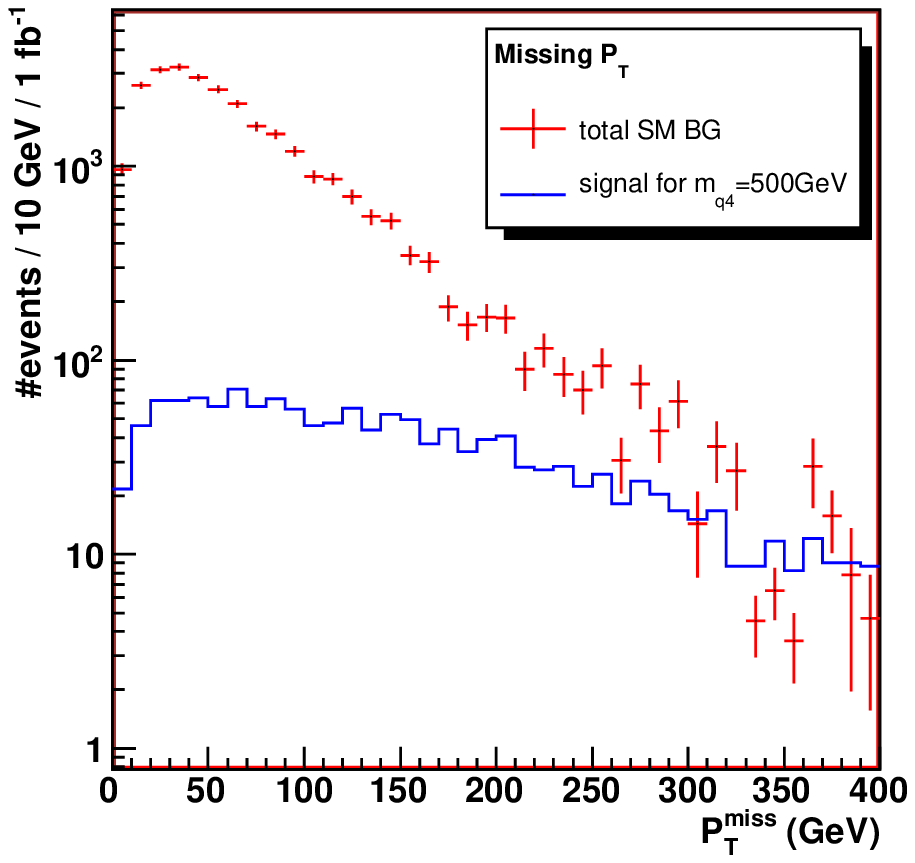}}\vspace{-0.4cm}

\par\end{centering}

\begin{centering}
\subfigure[]{\includegraphics[width=0.42\textwidth,height=4.8cm]{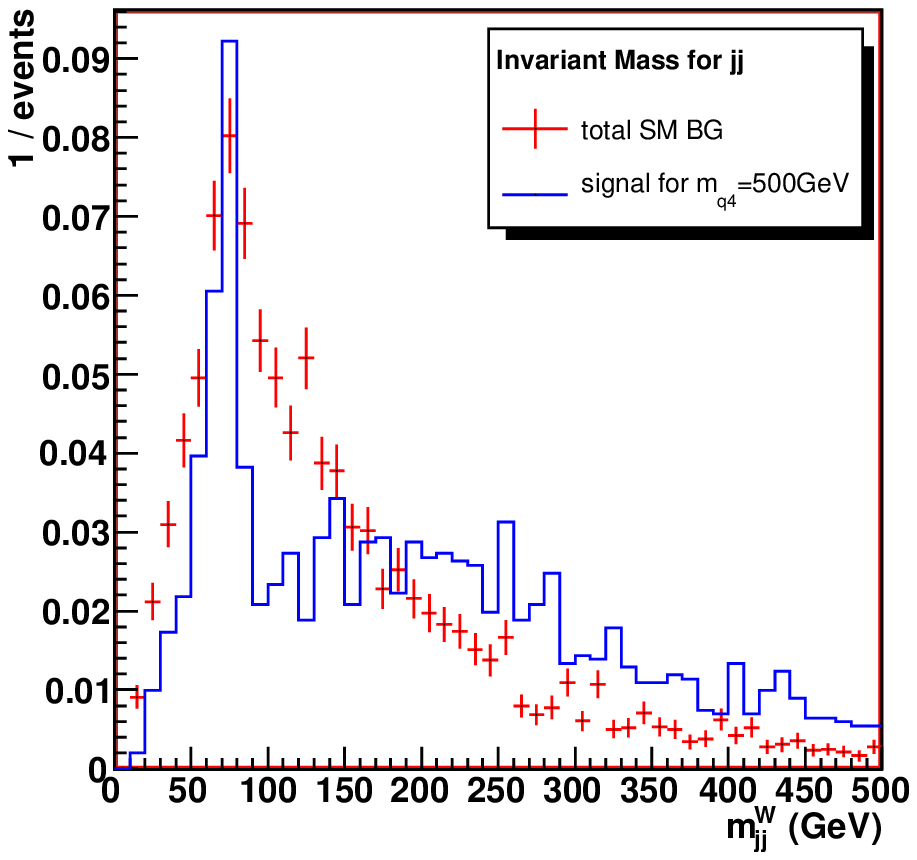}}\subfigure[]{\includegraphics[width=0.42\textwidth,height=4.8cm]{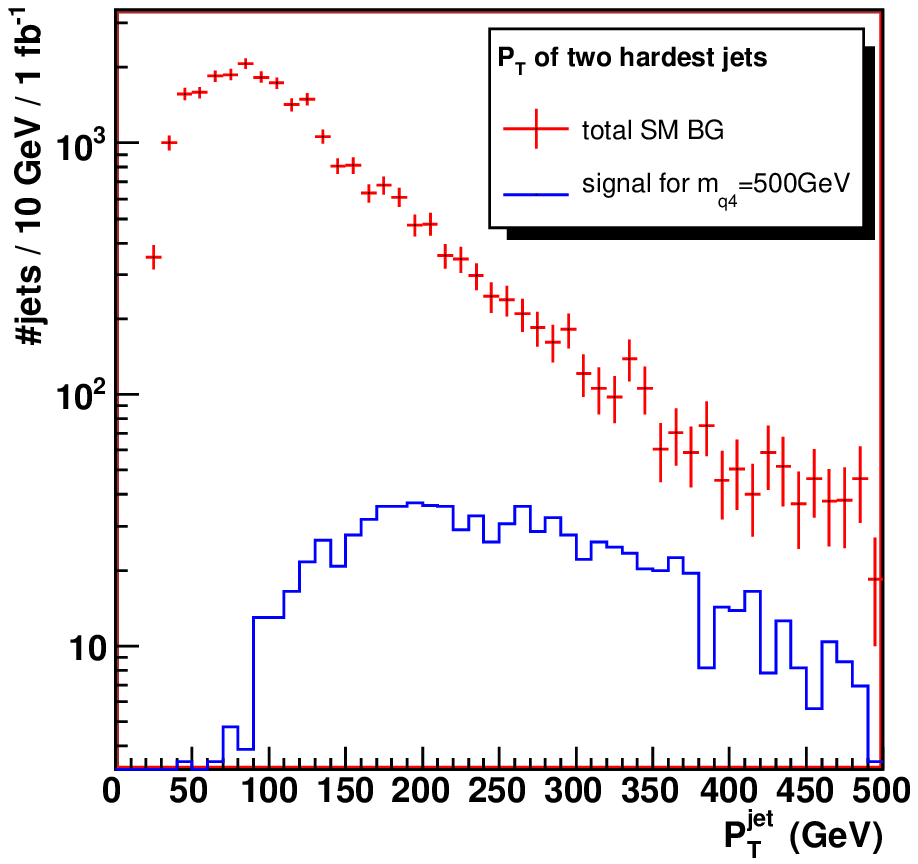}}
\par\end{centering}

\caption{Distributions of the kinematical observables for the signal events
(blue line) as compared to the backgrounds (red data points): (a)
transverse momentum for the highest-$p_{T}$ charged lepton, (b) missing
transverse momentum, (c) invariant mass of the hadronic $W$ candidates
reconstructed from two jets, and (d) transverse momentum for the two
hardest jets. The distributions in plots a, b, c and d are shown after
the application of the event selection requirements up to the criteria
number 1, 4, 5 and 6 in Table \ref{tab:Efficiencies}, respectively.
Signal and background histograms have been scaled to the same luminosity,
except in plot c, where the histograms have been normalized to unit
area.\label{fig:kinematic_distro}}

\end{figure}

The first step of the event selection is the requirement of a single
isolated lepton of transverse momentum, $\ptleptcut$, and at least
four jets with transverse momenta, $\ptjetcut$. The transverse momentum
of the highest-momentum isolated muon in each event is shown in Fig.
\ref{fig:kinematic_distro}a. The four highest-energy jets are required
not be $b$-tagged, as determined by ATLFASTB\cite{R-AtlFast}, a
fast $b$-tagging simulation program, which utilizes a $p_{T}$ dependent
parameterization of tagging efficiencies. For instance, at high momenta
($p_{T}^{jet}>100\gev$) the tagging efficiency for b, c and light
jets are 50\%, 7.6\% and 0.6\%, respectively.

The leptonically decaying $W$ boson is reconstructed by attributing
the total missing transverse momentum in the event, shown in Fig.
\ref{fig:kinematic_distro}b, to the lost neutrino, and using the
nominal mass of the $W$ as a constraint. The two-fold ambiguity in
the longitudinal direction of the neutrino is resolved by choosing
the solution with the lower neutrino energy. The four-momenta of the
third and fourth most energetic jets in the event are combined to
reconstruct the hadronically decaying $W$ boson. Due to the high
momentum of the $W$ boson in the signal events particularly for the
high values of the $q_{4}$ mass, the jets are not always resolved
in the detector. When this happens, one of the two jets used in the
combination is a random jet, which spuriously increases the invariant
mass, $\mWhad$, of the reconstructed $W$. Such cases cause a long
high-end tail in the invariant mass distribution for the signal as
shown in Fig. \ref{fig:kinematic_distro}c. In order to reduce their
adverse effect on the final $m_{q_{4}}$ distributions, events with
$\mWhad>200\gev$ are rejected, even though the comparison of the
distributions for the signal and the background would suggest that
a looser criterion would benefit the final statistical significance. 

\begin{table}
\caption{Efficiencies of the selection criteria, as applied in the order listed,
for the $m_{q_{4}}$=500 (750)$\gev$ signal and the largest component
of the SM background ($t\,\bar{{t}}\, j$). \label{tab:Efficiencies}\smallskip{}
}

\begin{centering}
\begin{tabular}{c|c|c|c}
\# & Criterion & $\epsilon$-Signal (\%) & $\epsilon$-Background (\%)\tabularnewline
\hline
\hline 
1 & Single $e/\mu$, $\ptleptcut$ & 32 (32) & 29\tabularnewline
\hline 
2 & At least 4 jets, $\ptjetcut$  & 86 (84) & 84\tabularnewline
\hline 
3 & $b$-tagging veto & 92 (90) & 33\tabularnewline
\hline 
4 & Possible neutrino solution & 75 (71) & 76\tabularnewline
\hline
5 & $m_{jj}^{W}<200\gev$ & 50 (44) & 75\tabularnewline
\hline
6 & 2 hardest jets, $\hardjetcut$ & 94 (98) & 35\tabularnewline
\hline
7 & $|\Delta m_{Wj}^{q_{4}}|<100\gev$ & 56 (49) & 50\tabularnewline
\hline
 & Total efficiency, $\epsilon_{all}$ & 5.0 (3.6) & 0.8\tabularnewline
\end{tabular}
\par\end{centering}
\end{table}

The surviving events are used to obtain the invariant mass of the
new quark. Each reconstructed $W$ is associated with one of the two
hardest jets, for which the minimum transverse momentum requirements
are tightened to $\hardjetcut$. As observed in Fig. \ref{fig:kinematic_distro}d,
this tighter requirement has no significant effect on the signal,
while substantially reducing the background. A tighter $p_{T}^{jet}$
selection would start to skew the final invariant mass distributions.
Therefore the lower value of 100 GeV was chosen so that the analysis
results could be safely interpreted for lower $q_{4}$ masses as well.
The $W$-jet association ambiguity is resolved by selecting the combination
which results in the smallest difference between the masses of the
two reconstructed $q_{4}$ quarks in the same event. If this mass
difference is more than 100$\gev$ for either combination, the event
is rejected. The summary of the event selection cuts and their efficiencies
for both signal and background events are listed in Table \ref{tab:Efficiencies}
for a quark mass of 500$\gev$. These selection criteria were not
optimized for the $m_{q_{4}}=750$GeV case to be safely pessimistic.
The results of the reconstruction for quark masses of 500$\gev$and
750$\gev$ are shown in Fig. \ref{fig:Reconstructed} together with
various backgrounds for integrated luminosities of 1 and 10~fb$^{-1}$
respectively. The bulk of the background in both cases is due to $t\,\bar{t}\: j$
events as discussed before.

\section{Results}

\begin{figure}
\begin{centering}
\includegraphics[scale=0.38]{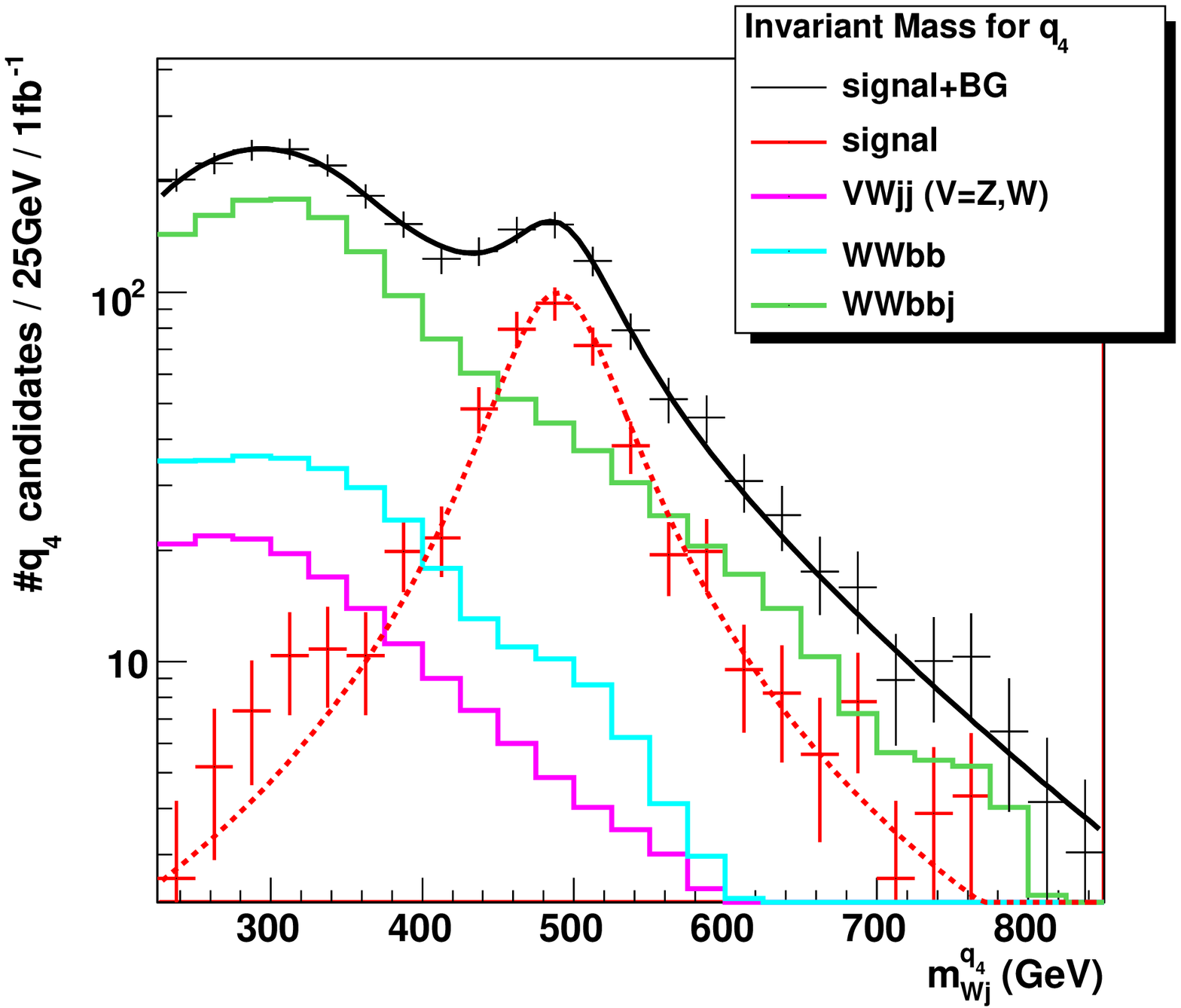}\includegraphics[scale=0.38]{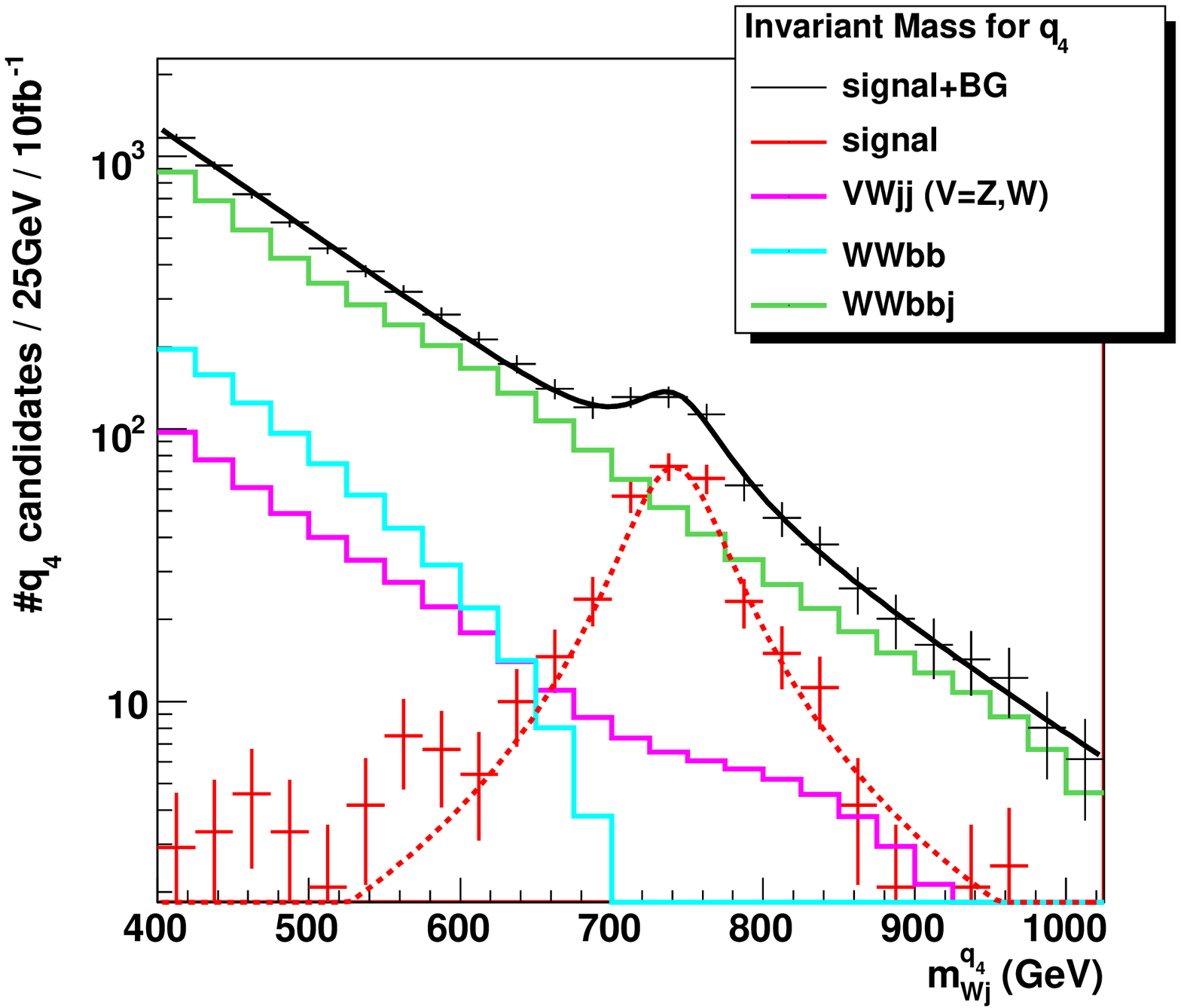}
\par\end{centering}

\caption{Invariant mass distributions for the reconstructed $q_{4}$ candidates
from signal and SM background events for a quark of mass 500$\gev$
(left) and 750$\gev$ (right). The histograms are populated by two
$q_{4}$ candidates per event. The colored solid lines show the backgrounds
from various processes, the solid black curve represents the fit to
the sum of the background and signal events. Also shown in red dashed
curve is the signal component of the fit.\label{fig:Reconstructed}}

\end{figure}

In order to extract the signal significance, an analytical function
consisting of a Crystal$~$Ball term \cite{R-xtal-ball} to represent
the background and a Breit-Wigner term to represent the signal resonance
is fitted to the total number of $q_{4}$ candidates in the invariant
mass plots of Fig. \ref{fig:Reconstructed}. In both plots, the fitted
function is shown in solid black, and its signal component is plotted
as a dashed red line. The shape of the background curve was also verified
against random fluctuations (as in Fig.~\ref{fig:Reconstructed}
left side in the 500-600~GeV region of the $WWbbj$ curve) by parameterizing
the background and then by generating a large sample of pseudoMC experiments.
It was found that with large statistics the Crystal$~$Ball is a very
accurate description of the background shape. The extracted number
of total signal events is in very good agreement with the actual number
of events in the signal Monte Carlo sample. The significance is estimated
as $S/\sqrt{S+B}$, where S(B) is the number of signal (background)
events determined from the Breit-Wigner (Crystal Ball) term of the
fitted function. As each event contributes two $q_{4}$ candidates
to the invariant mass histogram, the total number of signal (background)
events is obtained by taking half of the integral of the signal (background)
term within $\pm2\Gamma$ (twice the fullwidth at half maximum) of
the peak position of the signal. For the case of $m_{d_{4}}$=500$\gev$
(750$\gev$), with 1 fb$^{-1}$(10 fb$^{-1}$) of data, the signal
significance is found to be 9.2 (7.1). The number of events for these
two example cases for both signal and background are presented in
Table \ref{tab:The-event-count}.

\begin{table}
\caption{The expected number of signal and background events and the signal
significance for the two masses under consideration.}

\begin{centering}
\begin{tabular}{c|c|c}
 & 500$\gev$ & 750$\gev$ \tabularnewline
\hline
\hline 
Luminosity & 1 fb$^{-1}$ & 10 fb$^{-1}$\tabularnewline
\hline 
Signal & 192 & 134\tabularnewline
\hline 
Background & 244 & 226\tabularnewline
\hline 
$S/\sqrt{S+B}$ & 9.2 & 7.1\tabularnewline
\end{tabular}\label{tab:The-event-count}
\par\end{centering}
\end{table}

\section{Conclusion}

\begin{figure}
\begin{centering}
\includegraphics[bb=10bp 30bp 550bp 550bp,clip,scale=0.4]{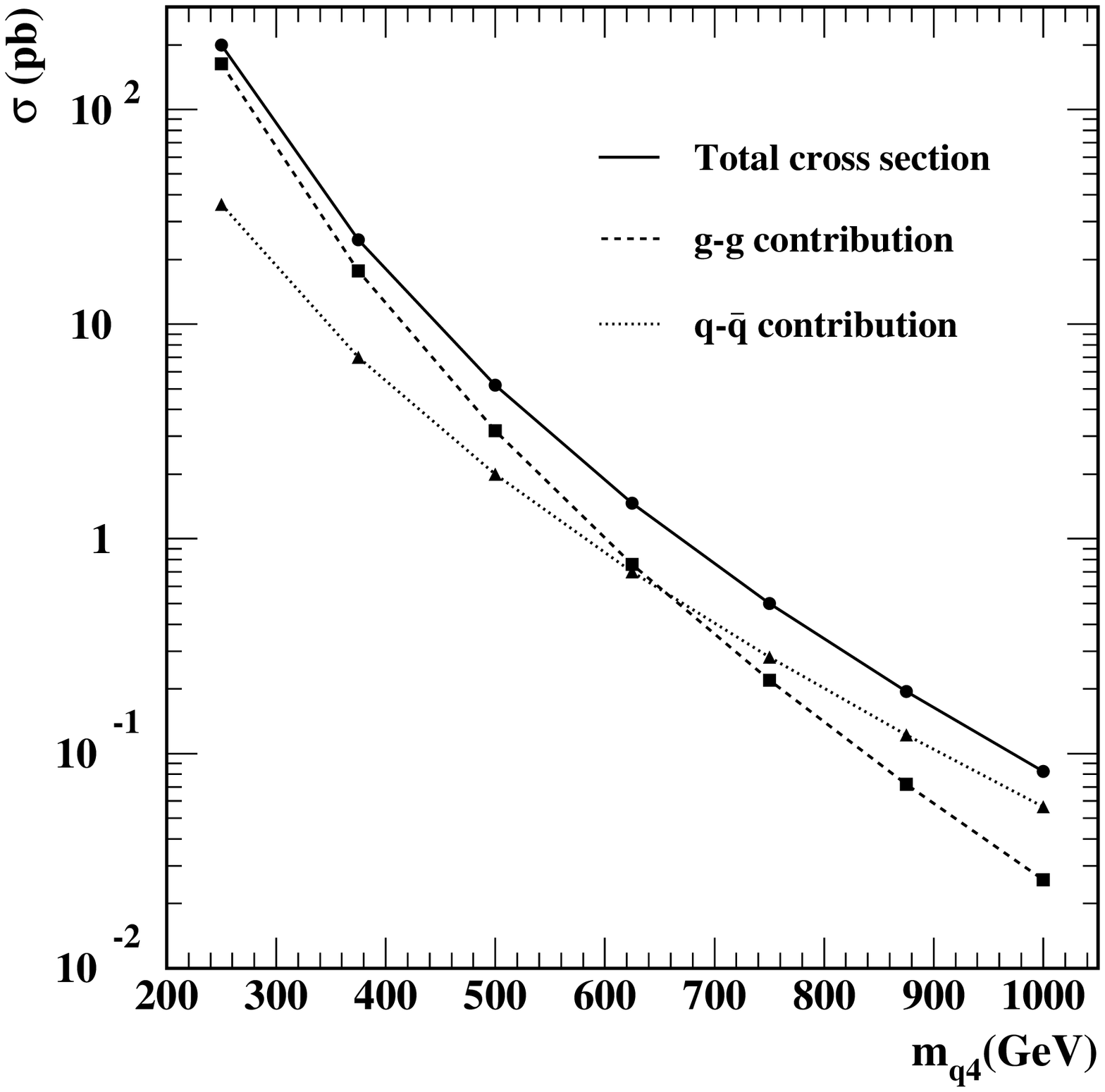}\includegraphics[bb=10bp 30bp 550bp 550bp,clip,scale=0.4]{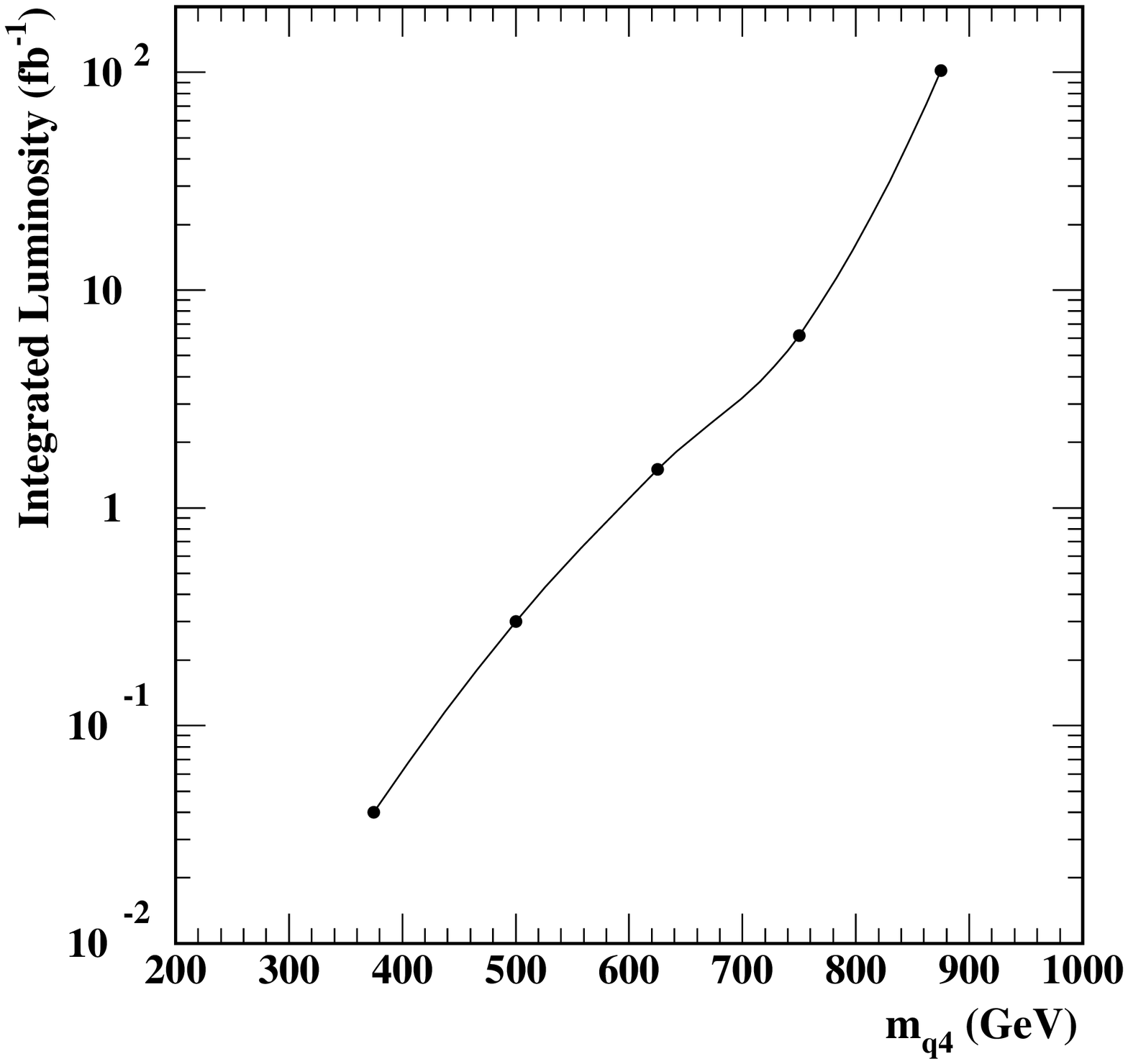}
\par\end{centering}

\caption{On the left, the $q_{4}\bar{q_{4}}$ pair production cross section
at the tree level and on the right, the integrated luminosity needed
for a $5\sigma$ discovery of the signal, both as a function of the
new quark mass. Only the pair production and the mixing to first two
families are considered.\label{fig:The-5-sigma}}

\end{figure}

The analysis can be extrapolated to other $q_{4}$ quark mass values
to estimate the amount of integrated luminosity necessary for a discovery.
Fig. \ref{fig:The-5-sigma} contains the fourth generation quark ($u_{4}$
and $d_{4}$ combined) pair production tree-level cross section showing
the contributions from gluon fusion and $q-\bar{q}$ annihilation.
For the selected parton distribution function, the latter becomes
more important at a quark mass of around 650$\gev$. The same figure,
on the right-hand side, shows the estimated integrated luminosity
required for 5$\sigma$ discovery as a function of the mass of the
new quark. The estimates on this plot are based on the cross sections
shown and the integration of the background function as obtained from
the fits presented in the analysis section. In all cases, the number
of signal events to be collected in order to reach the 5$\sigma$
significance is above 20. While this study is based on a fast simulation
of the detector response which was not fully validated and there are
uncertainties associated with the QCD scale, statistical errors etc,
we believe that the conservative selection cuts and the simplicity
of the reconstruction algorithms give reliability to the conclusions.

This study has shown that, if the fourth family quarks mix primarily
with the first two generations, a clear signal will be observed for
the mass range of interest within the first year of the low-luminosity
running at the LHC. On the other hand, if the mixing matrix is such
that the third SM family quarks play the dominant role, similar results
can be claimed for the $u_{4}$ quark, while the discovery of the
$d_{4}$ quark is likely to require more luminosity because of the
complexity of the event signature arising from the top-quark decays.
In either case, the first few years of the LHC data will resolve the
discussion on the possibility of four SM families within the context
of flavor democracy.

\subsection*{Acknowledgments}

The authors would like to thank Louis Tremblet and CERN Micro Club
for kindly providing computational facilities, Fabienne Ledroit and
Andy Parker for fruitful discussions. S.S. acknowledges the support
from the Turkish State Planning Committee under the contract DPT2006K-120470.
G.\"U.'s work is supported in part by U.S. Department of Energy Grant
DE FG0291ER40679. This work has been performed within the ATLAS Collaboration
with the help of the simulation framework and tools which are the
results of collaboration-wide efforts.

\end{document}